\documentclass[12pt]{article}
\usepackage{etex}
\usepackage{fullpage}
\usepackage{setspace}
\usepackage{comment}
\usepackage[pdftex]{graphicx}
\usepackage{rotating}
\usepackage{amsmath}
\usepackage{amsthm}
\usepackage{amssymb}
\usepackage{graphicx}
\usepackage{fullpage}
\usepackage{setspace}
\usepackage{natbib}
\usepackage{color}
\usepackage{booktabs}
\usepackage[linesnumbered, ruled]{algorithm2e}
\SetKwRepeat{Do}{do}{while}%

\usepackage{tikz, animate}
\usepackage{pgf}
\usepackage{pgfplots}
\usepackage{textpos}
\usepackage{graphicx}
\usepackage{pstricks}
\usepackage{framed}
\usepackage{varwidth}
\usepackage[normalem]{ulem}
\usepackage{xstring}
\usetikzlibrary{shapes}
\usetikzlibrary{arrows,decorations.pathmorphing,backgrounds,fit,positioning,shapes.symbols,chains}

\usepackage{breqn}

\newcommand{\T}{\intercal}

\newcommand{\pn}{\mathbb{P}_{n}}

\newcommand{\OMIT}[1]{\relax}   
\def\text{{\rm}}

 \newcommand{\bma}[1]{\mbox{\boldmath $#1$}}

 \newcommand{\bC}{ {\bma{C}} }

 \newcommand{\bH}{ {\bma{H}} }
 \newcommand{\bh}{ {\bma{h}} }

 \newcommand{\bv}{ {\bma{v}} }

 \newcommand{\bX}{ {\bma{X}} }

\newtheorem{thm}{Theorem}[section]
\newtheorem{lem}[thm]{Lemma}
\newtheorem{cor}[thm]{Corollary}
\theoremstyle{definition}

\newtheorem{rmrk}[thm]{Remark}

\begin{document}

\title{Functional feature construction for individualized treatment
regimes}
\author{Eric B. Laber\\ Department of Statistics\\ North Carolina State
University
  \and Ana-Maria Staicu \\ Department of Statistics\\ North Carolina
State University}


\pagestyle{empty}
\begin{center}
  \textbf{Sample Size Calculations for SMARTs} \\
  \textbf{Eric J. Rose$^{1}$, Eric B. Laber$^{1}$, 
    Marie Davidian$^{1}$, Anastasios A. Tsiatis$^{1}$, 
    Ying-Qi Zhao$^{2}$,
  Michael
    R. Kosorok$^{3}$}
  \\
$^1$Department of Statistics, North Carolina State University,
  Raleigh, NC, 27695, U.S.A.  \\
$^2$Public Health Sciences Division, Fred Hutchinson Cancer Research Center, 
Seattle, WA, 98109\\
$^3$Department of Biostatistics, University of North Carolina,
Chapel Hill, NC, 27599
\end{center}

\begin{abstract}\noindent
  Sequential Multiple Assignment Randomized Trials (SMARTs)
  are considered the gold standard for 
  estimation and evaluation of treatment regimes.  
SMARTs are
  typically sized to ensure sufficient power for a simple comparison,
  e.g., the comparison of two fixed 
 treatment sequences.  Estimation of an optimal
  treatment regime is conducted as part of a secondary and
  hypothesis-generating analysis with formal evaluation of the
  estimated optimal regime deferred to a follow-up trial.   
  However, running a follow-up trial to evaluate an estimated optimal
  treatment regime is costly and time-consuming; furthermore, the
  estimated optimal regime that is to be evaluated in such a follow-up
  trial may be far from optimal if the original trial was underpowered
  for estimation of an optimal regime.  We derive sample size
  procedures for a SMART that ensure: (i) sufficient power for
  comparing the optimal treatment regime with standard of care; 
  and (ii) the estimated optimal regime is
  within a given tolerance of the true optimal regime with
  high-probability.  
  We establish asymptotic validity of the proposed procedures 
  and demonstrate their finite sample performance in a series
  of simulation experiments.  
\end{abstract}

\pagebreak
\setcounter{page}{1}
\pagestyle{plain}
\section{Introduction}
A treatment regime is a sequence of functions, one per stage of
clinical intervention, that map up-to-date patient information to a
recommended treatment.  An optimal treatment regime maximizes the mean
of some cumulative clinical outcome when applied to select treatments
for individuals in a population of interest \citep[][]{murphyZThree,
  robinsTF}.  Thus, an optimal treatment regime leads to better
overall healthcare by adapting treatment to the evolving health status
of each patient; consequently, optimal treatment regimes have become a
primary means of operationalizing precision medicine.
Optimal treatment regimes have been estimated across a wide range
of application domains  including breastfeeding \citep[][]{moodie2012q},
bipolar disorder \citep[][]{wu2015will, listBasedRegimes}, cancer
\citep[][]{thall2000evaluating, zhao2011reinforcement, wang2012,
  baqun, zhang2015using, thallBayesQ}, 
cystic fibrosis \citep[][]{zhou2017residual}, diabetes
\citep[][]{ertefaie2014constructing, luckett2016estimating},
depression \citep[][]{yingqi}, HIV \citep[][]{moodie, van2005history,
  cain2010start}, smoking cessation
\citep[][]{chakraborty2009inference}, substance abuse
\citep[][]{nahum2017smart}  among others.

Sequential Randomized Multiple Assignment Randomized Trials
\citep[SMARTs][]{lavori2000design, lavori2004dynamic,
  murphy2005experimental, kidwell2014smart} are the gold standard for
estimating and evaluating treatment regimes
\citep[][]{murphy2007developing, lei2012smart, bibhasBook,
  kosorok2015adaptive} and are increasingly common design in clinical
and intervention science \citep[][]{methCenterURL, methCenterURLNIH}.
However, sample size calculations for SMARTs are typically based on
power calculations for simple comparisons, e.g., comparison of the
mean outcome across two pre-specified treatment sequences
\citep[][]{murphy2005experimental, lei2012smart} and estimation of an
optimal treatment regime from data collected in a SMART are almost
always (we are not aware of an exception) conducted as part of
exploratory, hypothesis-generating analyses.   This approach
is aligned with the {\em estimate-and-validate} paradigm wherein: 
(i)  an optimal treatment regime is estimated using data collected in
a SMART; and (ii)  the performance of the estimated optimal
regime is validated in a follow-up trial where the estimated
regime is compared head-to-head with standard of care
\citep[][]{murphy2005experimental}.    This approach is appealing
in that it avoids a number of nontrivial technical issues
associated with estimating and evaluating a treatment regime
using the same data \citep[][]{robinsTF, moodieT, bibhas, song,
chakraborty2014inference, laber2014statistical}; furthermore, sample
size formulae for the comparison of fixed treatment sequences or other
commonly used criteria to size SMARTs
are straightforward in that they resemble those commonly used in
non-sequential randomized trials.       Another reason that
sample size calculations for SMARTs are often based on simple comparisons is the
seemingly widely held belief
that sizing a trial to guarantee
frequentist
operating characteristics for an estimated optimal regime, e.g.,
providing a performance guarantee for the estimated regime or powering
a comparison of the performance of the optimal regime with standard of
care, would either: (i) require a prohibitively large sample
size; and/or (ii)  rely on unrealistic assumptions about the underlying
data-generating model.    We provide evidence that for a for a large class of
generative models  neither of these beliefs 
appear to be well-founded.  

We derive sample size procedures for a SMART that ensure sufficient
power in comparing the mean outcome under the optimal regime with
standard of care and that the estimated optimal regime will be within
a given tolerance of the optimal policy with a given probability.  
The proposed sample sized procedures we develop here are at two possible
extremes in terms of modeling assumptions.   Our first procedure 
imposes significant parametric structure on the
data-generating model and consequently we are able to derive sample
size formulae that resemble the comparison of two means and require 
elicitation or estimation of a single scalar parameter.  Our second
procedure imposes structure only on moments and tail behavior of 
some components of the data generating model and then uses the bootstrap
with oversampling to estimate a sufficient sample size; as this
procedure imposes less structure the resulting estimator is more
variable and consequently the estimated sample size tends to be
larger.   One reason for considering these two extremes is that they
provide a basis for intermediate procedures that impose as much
structure, as appropriate for a given application. We leave such 
intermediate approaches to future work.   

We perceive the proposed work as making the following contributions:
(i) it provides the first rigorous yet practical sample size procedures for
estimating and evaluating optimal treatment regimes using SMARTs;  
(ii) it generates new knowledge about how much additional data would be
needed to estimate a high-quality regime and consequently a provides
a sense of how `underpowered' existing SMARTs are for estimating
optimal treatment regimes; and (iii) it provides theoretical
guarantees for bootstrap oversampling for sample size calculations 
that are of independent interest.    Furthermore, the proposed
criteria used to derive our samples size procedures are closely
related to those used in \citet{laber2015using} to size a single stage two-arm
trial to estimate an optimal regime, however, the current procedure
applies
to multistage trials and provides
considerably stronger performance guarantees for the estimated optimal
regime.  

In Section 2, we provide the setup and notation.  In Section 3, we
derive our sample size procedures and state their theoretical
properties.  In Section 4,  we evaluate the finite sample performance
of the proposed sample size procedures in a series of simulation
experiments.  A discussion of the proposed methodology and open
problems is provided in Section 5.

\section{Setup and notation}
We consider choosing the sample size, $n$, for a two-stage SMART that
will produce data,
$\mathcal{D}_{n} = \left\lbrace (\bX_{1,i}, A_{1,i}, \bX_{2,i},
  A_{2,i}, Y_i)\right\rbrace_{i=1}^{n}$,
which comprises $i.i.d.$ trajectories of the form
$(\bX_1, A_1, \allowbreak\bX_2, \allowbreak A_2, Y)$ where:
$\bX_1\in\mathbb{R}^{p_1}$ denotes baseline subject information;
$A_1\in\left\lbrace -1,1\right\rbrace$ denotes the first assigned
treatment; $\bX_2\in\mathbb{R}^{p_2}$ denotes subject information
collected during the course of the first treatment;
$A_2 \in\left\lbrace -1,1\right\rbrace$ denotes the second assigned
treatment; and $Y\in\mathbb{R}$ denotes the outcome, coded so that
higher is better.  Define $\bH_1 = \bX_1$ and
$\bH_2 = (\bX_1^{\T}, A_1, \bX_2^{\T})^{\T}$ so that $\bH_t$ denotes the
history at time $t=1,2$. 
For simplicity, we assume that the trial will employ simple one-to-one
randomization so that $P(A_t=a_t\big|\bH_t) = 1/2$ with probability
one
for $a_t \in \left\lbrace -1,1\right\rbrace,\,t=1,2$; extensions to
more complex randomization schemes including those with feasible sets
of treatments is straightforward \citep[][]{schulte}.  
 A treatment regime in this context is a pair
of functions $\pmb{\pi} = (\pi_1, \pi_2)$ where
$\pi_t:\mathrm{dom}\,\bH_t\rightarrow \mathrm{dom}\,A_t$ so that a
decision maker following $\pmb{\pi}$ would recommend treatment
$\pi_t(\bh_t)$ to a patient presenting with $\bH_t = \bh_t$ at time
$t=1,2$.    We define a treatment regime as optimal if it leads to maximal 
mean outcome if applied to the population from which $\mathcal{D}_n$
is drawn 
\citep[other definitions of optimality are possible, see][]{
kosorok2015adaptive, linnQIQ}.   To formally define
an optimal treatment regime, we use 
 potential outcomes \citep[][]{rubin, splawa1990application}.

Let $\bH_2^*(a_1)$ denote the potential second history under
initial treatment $a_1$ and $Y^*(a_1, a_2)$ the potential
outcome under treatment sequence $(a_1, a_2)$.  The
potential outcome under a regime $\pmb{\pi}$ is 
\begin{equation*}
Y^*(\pmb{\pi}) = \sum_{(a_1,a_2)}Y^*(a_1,a_2)1_{\pi_1(\bH_1) = a_1}
1_{\pi_2\lbrace \bH_2^*(a_1)\rbrace = a_2},
\end{equation*}
where $1_u$ is an indicator that $u$ is true.  
For any regime, $\pmb{\pi}$, define
$V(\pmb{\pi}) = \mathbb{E}Y^*(\pmb{\pi})$;
an optimal regime,
$\pmb{\pi}^{\mathrm{opt}}$, satisfies
$V(\pmb{\pi}^{\mathrm{opt}}) \ge V(\pmb{\pi})$
for all $\pmb{\pi}$.   Our sample size procedures depend on an
estimator of $\pmb{\pi}^{\mathrm{opt}}$, in order to construct such
an estimator,  we make the following assumptions:
(C1) sequential ignorability, $\left\lbrace 
\bH_2^*(a_1), Y^*(a_1, a_2)\,:\, (a_1, a_2) \in \left\lbrace -1,1\right\rbrace^{2}
\right\rbrace \perp A_t \big| \bH_t$ for $t=1,2$;  
(C2) positivity, $P(A_t=a_t|\bH_t) > 0$ with probability one for each
$a_t \in \left\lbrace -1, 1\right\rbrace$ for $t=1,2$; and
(C3)  consistency, $Y = Y^*(A_1, A_2)$ and $\bH_2 = \bH_2^*(A_1)$. 
These assumptions are standard in the context of estimating
optimal treatment regimes \citep[][]{robinsTF, schulte, bibhasBook} with 
(C1) and (C2) holding by design in a SMART.  

Under these assumptions, the optimal regime can be characterized in
terms of the data-generating model as follows.  Define
$Q_{2}(\bh_2, a_2) = \mathbb{E}(Y|\bH_2=\bh_2, A_2=a_2)$ and
$Q_{1}(\bh_1, a_1) = \mathbb{E}\left\lbrace \max_{a_2}Q_2(\bH_2, a_2)
  \big|\bH_1=\bh_1, A_1=a_1\right\rbrace$,
then $\pi_{t}^{\mathrm{opt}}(\bh_t) = \arg\max_{a_t}Q_{t}(h_t, a_t)$
for $t=1,2$ \citep[see][]{murphyZFive, schulte}.  Furthermore, it can
be seen that
$V(\pmb{\pi}^{\mathrm{opt}}) = \mathbb{E}\max_{a_1}
Q_1(\bH_1, a_1)$.
Our sample size procedures are based on constructing estimators of
$Q_t(\bh_1, a_t)$ for $t=1,2$ and subsequently deriving plug-in estimators of
$\pmb{\pi}$; these  procedures vary in the structure we impose on 
these functions.  Before describing specific estimators, we state
properties of these estimators that we would like to ensure hold with 
high-probability provided the sample size is sufficiently large. 

Let $\widehat{\pmb{\pi}}_{n}$ denote an estimator of 
$\pmb{\pi}^{\mathrm{opt}}$ and let $B_0 > 0$, 
$\gamma, \alpha,\eta, \epsilon,\zeta \in
(0,1)$ be constants.  Our goal is to choose $n$ so that:
\begin{itemize}
  \item[(POW)] there exists an $\alpha$-level test of $H_0:
    V(\pmb{\pi}^{\mathrm{opt}})  \le B_0$ based on
    $\widehat{\pmb{\pi}}_{n}$ that has power at least
    $(1-\gamma)\times 100 + o(1)$
    provided
    $V(\pmb{\pi}^{\mathrm{opt}}) \ge B_0 + \eta$;
\item[(OPT)]  $P\left[
    \mathbb{E}\left\lbrace
      Y^*(\widehat{\pmb{\pi}}_{n})\big|\mathcal{D}_{n}
      \right\rbrace \ge 
    V(\pmb{\pi}^{\mathrm{opt}}) - \epsilon
    \right] \ge 1-\zeta + o(1)$.
\end{itemize}
Condition (POW) ensures sufficient power to test the effectiveness
of the optimal treatment regime relative to some baseline expected
outcome, $B_0$, e.g., the expected outcome under some standard
of care.  Condition (OPT) ensures that the expected performance of the
estimated optimal regime will be near-optimal with high-probability.  
These conditions are analogous to those used to size a one-stage
clinical trial for estimation of an optimal regime
except that (OPT) controls the
performance of the estimated optimal regime whereas 
\citet{laber2015using} control the {\em estimated performance} of the 
estimated optimal regime.   
  Furthermore, like the one-stage
setting, our sample size procedures depend on approximating the
sampling distribution of an estimator of
 $\mathbb{E}Y^*(\pmb{\pi}^{\mathrm{opt}})$; however, as we will later
illustrate, 
constructing a high-quality approximation is markedly more complex
in the multistage setting \citep[see also][]{dawid1994selection, chakraborty2009inference,
  moodieT, hirano2012, chakraborty2014inference, 
laber2014statistical, luedtke2016statistical}.  

\section{Sample size procedures}
We derive two sample size procedures. The first procedure imposes
more parametric structure on  the joint distribution of 
$(\bX_1, A_1, \bX_2, A_2, Y)$ than is typical in $Q$-learning 
and thereby avoids (or rather assumes away) some of the complexities
associated with non-regularity and exceptional laws
\citep[][]{robinsTF, chakraborty2009inference, moodieT, mofn,
  chakraborty2014inference,
 laber2014statistical, song2015penalized}. The second
proposed procedure does not impose as much
parametric structure but at the expense of a more
complex and potentially conservative sample size estimator.  
\subsection{Normality-based sample size procedure}
We make the following assumptions about the generative
model:
\begin{itemize}
  \item[(AN1)] $Q_2(\bh_2, a_2) = 
    \bh_{2,0}^{\T}\beta_{2,0}^* + a_2\bh_{2,1}^{\T}\beta_{2,1}^*$,
    where $\bh_{2,0}\in\mathbb{R}^{p_{2,0}},
    \bh_{2,1}\in\mathbb{R}^{p_{2,1}}$ 
    are summaries of $\bh_2$ and
    $\beta_{2,0}^*\in\mathbb{R}^{p_{2,0}},
      \beta_{2,1}^*\in \mathbb{R}^{p_{2,1}}$ 
    are unknown
    parameters;
 \item[(AN2)]
   $\mathbb{E}\left(\bH_{2,0}^{\T}\beta_{2,0}^*\big|\bH_1=\bh_1, 
     A_1=a_1\right) = \bh_{1,0}^{\T}\xi_{1,0}^* + a_1
   \bh_{1,1}^{\T}\xi_{1,1}^*$,
   where $\bh_{1,0}\in\mathbb{R}^{p_{1,0}},
   \bh_{1,1}\in\mathbb{R}^{p_{1,1}}$ 
   are summaries of $\bh_1$ and 
   $\xi_{1,0}^*\in\mathbb{R}^{p_{1,0}},
   \xi_{1,1}^*\in\mathbb{R}^{p_{1,1}}$ 
   are unknown parameters; 
 \item[(AN3)] $\bH_{2,1}^{\T}\beta_{2,1}^* = \bH_{1,2}^{\T}\varpi_{1,2}^* 
   +A_1\bH_{1,3}^{\T}\varpi_{1,3}^* + \tau^* Z$,
   where $\bH_{1,2}\in\mathbb{R}^{p_{1,2}},
   \bH_{1,3}\in\mathbb{R}^{p_{1,3}}$ 
   are summaries of $\bH_1$, 
   $Z$ is a standard normal random variable which is independent of
   $\bH_1, A_1$, and
   $\tau^* > 0$, $\varpi_{1,2}^*\in\mathbb{R}^{p_{1,2}}, 
   \varpi_{1,3}^*\in\mathbb{R}^{p_{1,3}}$, are unknown
   parameters;
\item[(AN4)] 
$\left(
\bH_{1,0}^{\T}\xi_{1,0}^*,
        \bH_{1,1}^{\T}\xi_{1,1}^*,
        \bH_{1,2}^{\T}\varpi_{1,2}^*, 
        \bH_{1,3}^{\T}\varpi_{1,3}^*\right)^{\T} \sim
      \mathrm{Normal}(\omega^*, \Omega^*)$, 
      where $\omega^* \in\mathbb{R}^4$ and $\Omega^* \in
      \mathbb{R}^{4\times 4}$ are unknown parameters.  
\end{itemize}
Assumptions (AN1)-(AN3) are similar to those used in interactive
$Q$-learning \citep[IQ-learning][]{iqLearnSlagathor} except that
in IQ-learning, (AN3) is replaced with a more general location-scale model than 
the normal linear model used here.   The summaries of the history
$\bh_t$ for $t=1,2$ can include basis expansions or other non-linear
terms as needed.   These assumptions were motivated
by a desire to create a generative model that is conceptually
consistent with the analysis model used in linear Q-learning which
remains the most commonly used method for estimating an optimal
treatment regime from SMARTs.  
 Assumption (AN4) is not 
required by IQ-learning as it conditions on 
$\bH_1$.  The assumption of joint normality could be relaxed, 
for example, by using a copula or semi-parametric model, but at the
expense of more complex expressions that are less amenable to 
sample size calculations.   Therefore we
shall not consider such generalizations further.

The following results, which are proved in the Supplemental Materials,
will be used to inform the construction of an estimator of the optimal
treatment regime \cite[see][for related expressions]{schulte,
  iqLearnSlagathor}.    Let $\Phi$ denote the cumulative distribution
function of a standard normal random variable.  
\begin{lem}\label{lemStefanski} 
Assume (AN1)-(AN3).  For any $\beta_1 =
(\xi_{1,0}^{\T},\xi_{1,1}^{\T},
\varpi_{1,2}^{\T},\varpi_{1,3}^{\T})^{\T}$ define 
\begin{multline*}
Q_1(\bh_1, a_1;\beta_1, \tau) = 
\bh_{1,0}^{\T}\beta_{1,0} + a_1\bh_{1,1}^{\T}\beta_{1,1}+
\frac{2\tau}{\sqrt{2\pi}}\exp\left\lbrace
-\frac{
\left(
\bh_{1,2}^{\T}\beta_{1,2} + a_1\bh_{1,3}^{\T}\beta_{1,3}
\right)^2
}{
2\tau^2
} 
\right\rbrace
\\ +
\left(
\bh_{1,2}^{\T}\beta_{1,2} + a_1\bh_{1,3}^{\T}\beta_{1,3}
\right)
\left[
1- 2\Phi
\left\lbrace
-\frac{
\left(
\bh_{1,2}^{\T}\beta_{1,2} + a_1\bh_{1,3}^{\T}\beta_{1,3}
\right)
}{
\tau
}
\right\rbrace
\right].
\end{multline*}
Then, $Q_1(\bh_1, a_1) = Q_1(\bh_1, a_1;\beta_1^*, \tau^*)$. 
\end{lem}\noindent
Let
$W(\bH_1, \beta_1)  = (\bH_{1,0}^{\T}\xi_{1,0}, 
\bH_{1,1}^{\T}\xi_{1,1}, \bH_{1,2}^{\T}\varpi_{1,2},
\bH_{1,3}^{\T}\varpi_{1,3})^{\T}$ and define
$g:\mathbb{R}^4\rightarrow \mathbb{R}$ as
\begin{equation*}
g(\bv) = \max_{\rho\in\left\lbrace -1,1\right\rbrace}\left(
v_1 + \rho v_2 + \frac{1}{\sqrt{2\pi}}\exp\left\lbrace
-\frac{
(v_3 + \rho v_4)^2
}{
2
}
\right\rbrace
+ (v_3 + \rho v_4)
\left[
1-2\Phi\lbrace-\left(
v_3 + \rho v_4
\right)\rbrace
\right]
\right);
\end{equation*}
it follows from Lemma (\ref{lemStefanski}) that 
$\max_{a_1}Q_{1}(\bH_1, a_1) = \tau^*g\left\lbrace
W(\bH_1, \beta_1^*)/\tau^*
\right\rbrace$.  
Let $\psi(\bv; \omega, \Omega)$ denote the density of 
a multivariate normal distribution with mean $\omega \in \mathbb{R}^4$ 
and covariance $\Omega \in \mathbb{R}^{4\times 4}$ and write
$\mathrm{vech}(\Sigma)$ to denote the vector-half operator
of symmetric matrix $\Sigma$ \citep[][]{henderson1979vec}.  
The following result shows
that $V(\pmb{\pi}^{\mathrm{opt}})$ is a smooth function of
$\omega^*$, $\tau^*$, and $\Omega^*$.  
\begin{cor}
Assume (AN1)-(AN4) and let $g:\mathbb{R}^4\rightarrow \mathbb{R}$ be
defined as above.  Then
\begin{equation*}
V(\pmb{\pi}^{\mathrm{opt}}) = \nu\left\lbrace
\tau^*, \omega^*, \mathrm{vech}(\Omega^*)
\right\rbrace = 
\int_{\mathbb{R}^4} \tau^*g\left(\bv/\tau^*\right) 
\psi(\bv; \omega^*, \Omega^*)d\bv.
\end{equation*}
\end{cor}\noindent
Thus, given estimators $\widehat{\tau}_{n}$, $\widehat{\omega}_{n}$, and
$\widehat{\Omega}_{n}$ of $\tau^*$, $\omega^*$, and $\Omega^*$, one can
use the preceding result to construct a plugin estimator of 
$V(\pmb{\pi}^{\mathrm{opt}})$.    We next describe how to 
construct these estimators.  

Let $\pn$ denote the empirical measure. 
Define $Q_2(\bh_2, a_2;\beta_2) = \bh_{2,0}^{\T}\beta_{2,0} + 
a_2\bh_{2,1}^{\T}\beta_{2,1}$ and subsequently define 
$\widehat{\beta}_{2,n} = \arg\min_{\beta_2}
\pn \left\lbrace Y- Q_2(\bH_2, A_2;\beta_2)\right\rbrace^2$.  
Thus, $\widehat{Q}_{2,n}(\bh_2, a_2) = Q_2(\bh_2, a_2;
\widehat{\beta}_{2,n})$
and 
the estimated optimal rule at the second stage is
$\widehat{\pi}_{2,n}(\bh_2) = \arg\max_{a_2}\widehat{Q}_{2,n}(\bh_2,
a_2)$.    Define 
\begin{eqnarray*}
\widehat{\xi}_{1,0,n}, \widehat{\xi}_{1,1,n}  &=&
\arg\min_{\xi_{1,0}, \xi_{1,1}}\pn\left(
\bH_{2,0}^{\T}\widehat{\beta}_{2,0,n} - 
\bH_{1,0}^{\T}\xi_{1,0} - A_1\bH_{1,1}^{\T}\xi_{1,1}
\right)^2 \\ 
\widehat{\varpi}_{1,2,n}, \widehat{\varpi}_{1,3,n} &=&
                                                     \arg\min_{\varpi_{1,2},\varpi_{1,3}}
\pn\left(
\bH_{2,1}^{\T}\widehat{\beta}_{2,1,n} - \bH_{1,2}^{\T}\varpi_{1,2} - 
A_1\bH_{1,3}^{\T}\varpi_{1,3}
\right)^2
\end{eqnarray*}
so that $\widehat{\beta}_{1,n} = (\widehat{\xi}_{1,0,n}^{\T},
\widehat{\xi}_{1,1,n}^{\T}, \widehat{\varpi}_{1,2,n}^{\T}, 
\widehat{\varpi}_{1,3,n}^{\T})^{\T}$ 
is the least-squares estimator of $\beta_1^* = 
\big(\xi_{1,0}^{*\T},\xi_{1,1}^{*\T},\allowbreak \varpi_{1,2}^{*\T},
\varpi_{1,3}^{*\T}\big)^{\T}$.  In addition, define
$\widehat{\tau}_{n}^2 = \pn\left\lbrace
\bH_{2,1}^{\T}\widehat{\beta}_{2,1} -
\bH_{1,2}^{\T}\widehat{\varpi}_{1,2,n}
-A_1\bH_{1,3}^{\T}\widehat{\varpi}_{1,3,n}
\right\rbrace^2$.  
The plugin estimator of $Q_1(\bh_1, a_1)$,
based on (\ref{lemStefanski}), is  $\widehat{Q}_{1,n}(\bh_1, a_1) = Q_1(\bh_1,
a_1;\widehat{\beta}_{1,n},
\widehat{\tau}_{n})$ and the estimated optimal
decision rule at the first stage is
$\widehat{\pi}_{1,n}(\bh_1) = \arg\max_{a_1}
\widehat{Q}_{1,n}(\bh_1, a_1)$.  
Furthermore,  define
$\widehat{\omega}_{n} = \pn W(\bH_1, \widehat{\beta}_{1,n})$  and
$\widehat{\Omega}_{n} = \pn \left\lbrace W(\bH_1, \widehat{\beta}_{1,n}) -
  \widehat{\omega}_{n}
\right\rbrace\left\lbrace W(\bH_1, \widehat{\beta}_{1,n}) - 
\widehat{\omega}_{n}\right\rbrace^{\T}$.   The plugin
estimator of $V(\pmb{\pi}^{\mathrm{opt}})$ is  
\begin{equation*}
\widehat{V}_{n} = \nu\left\lbrace 
\widehat{\tau}_{n}, \widehat{\omega}_{n}, \mathrm{vech}\left(
\widehat{\Omega}_{n}\right)
\right\rbrace
= 
\int_{\mathbb{R}^4} \widehat{\tau}_{n} 
g\left(\bv/\widehat{\tau}_{n}\right) 
\psi(\bv; \widehat{\omega}_{n}, \widehat{\Omega}_{n})d\bv.
\end{equation*}
To establish consistency and asymptotic normality of
$\widehat{V}_{n}$ we assume:
\begin{itemize}
\item[(AN5)] $\left\lbrace 
\tau^*, \omega^{*\T}, \mathrm{vech}(\Omega^*)\right\rbrace^{\T} \in 
\Theta \subseteq \mathbb{R}^{15}$, where $\Theta$ is compact; 
\item[(AN6)] $\sqrt{n}\left[
\left\lbrace \widehat{\tau}_{n}, \widehat{\omega}_{n}^{\T},
\mathrm{vech}(\widehat{\Omega}_{n})^{\T}\right\rbrace^{\T}
- \left\lbrace {\tau}^*, {\omega}^{*\T},
\mathrm{vech}({\Omega}^*)^{\T}\right\rbrace^{\T}
\right]
\leadsto \mathrm{Normal}(0, \Sigma^*)$, where
$\Sigma^*\in\mathbb{R}^{15\times 15}$ is positive definite.  
\end{itemize}
Condition (AN6) follows from moment conditions that are common
in $M$-estimation; we provide sufficient conditions for (AN6) 
in the Supplemental Materials.  
\begin{lem}
Assume (C1)-(C3) and  (AN1)-(AN6).  Then,
\begin{equation*}
\sqrt{n}\left\lbrace
\widehat{V}_{n} - V(\pmb{\pi}^{\mathrm{opt}})
\right\rbrace  \leadsto \mathrm{Normal}\left(
0, \sigma^{*2}\right),
\end{equation*}
where $\sigma^{*2}
= \nabla \nu\left\lbrace
\tau^*, \omega^*, \mathrm{vech}(\Omega^*)
\right\rbrace^{\T} \Sigma \nabla \nu\left\lbrace
\tau^*, \omega^*, \mathrm{vech}(\Omega^*)
\right\rbrace$.  
\end{lem}
Let $\widehat{\sigma}_{n}^2$ be a consistent estimator of 
$\sigma^{*2}$ and let $z_{1-\varrho}$ the $(1-\varrho)$ quantile of a
standard normal distribution, then 
a test that rejects when 
$\sqrt{n}\left\lbrace \widehat{V}_{n} -
  B_0\right\rbrace/\widehat{\sigma}_{n} 
\ge z_{1-\alpha}$ is an (asymptotic) $\alpha$-level test 
of $H_0:V(\pmb{\pi}^{\mathrm{opt}}) \le B_0$ with power exceeding 
$\Phi(z_{\alpha} + \sqrt{n}\eta/\sigma^*) + o(1)$ when
$V(\pmb{\pi}^{\mathrm{opt}}) \ge B_0 + \eta$.  Thus, choosing
$n = \left\lceil ({\sigma}^{*2}/\eta^2) \left\lbrace \Phi^{-1}(1-\gamma)
    + z_{1-\alpha}\right\rbrace^2\right\rceil$, satisfies (POW)
asymptotically.  This expression depends on $\sigma^*$, which is 
unknown in general, thus, a value for $\sigma^*$ must be elicited
from domain experts or estimated from historical data.  

The preceding sample size has a familiar form which is
unsurprising as it is 
derived from a test statistic which is asymptotically normal.  
However, what is perhaps more surprising,
is that a similar sample size formula can also be used to ensure
that condition (OPT) holds under the following regularity
conditions.
Define $\Delta Q_j = Q_j(\bH_j, 1) - Q_j(\bH_j, -1)$ 
and $\Delta \widehat{Q}_{j,n} = \widehat{Q}_{j,n}(\bH_j, 1)
-\widehat{Q}_{j,n}(\bH_j,-1)$ for $j=1,2$. 
To select $n$ so that (OPT) also holds we further assume:
\begin{itemize}
  \item[(AN7)] there exists positive sequences
    $\left\lbrace c_{n,j} \right\rbrace_{n\ge 1}$ and
    $\left\lbrace \ell_{n,j}\right\rbrace_{n\ge 1}$ satisfying 
    $\lim\inf_{n\rightarrow\infty}c_{n,j} \ge c_{0,j}>0$ and
    $\lim\inf_{n\rightarrow\infty}\ell_{n,j} \ge \ell_{0,j}> 0$ such
    that
    \begin{equation*}
      P\left\lbrace
        \sqrt{n}\big|
        \Delta \widehat{Q}_{j} - \Delta Q_{j}
        \big| > t 
        \right\rbrace \le \exp\left(
          -c_{n,j}t^{\ell_{n,j}}
          \right),
    \end{equation*}
    for all $n$ and $j=1,2$;  
\item[(AN8)] there exists $M_j, \kappa_j > 0$ such that 
$P\left( |\Delta Q_j| \le \epsilon 
\right) \le M_j\epsilon^{\kappa_j}$ for $j=1,2$ as $\epsilon \rightarrow 0$. 
\end{itemize}
The preceding assumptions are relatively mild with (AN7) being 
weaker than requiring a subexponetial tail; e.g., (AN7) and
(AN8) would be satisfied if the histories and outcomes are normally
distributed.  The following result characterizes the concentration 
of the marginal mean outcome under $\widehat{\pmb{\pi}}_{n}$ about
$\pmb{\pi}^{\mathrm{opt}}$ which can subsequently be used to choose
a sample size $n$ that satisfies (OPT).  
\begin{lem}
Assume (C1)-(C3) and (AN1)-(AN8).  Then there exists
$K$ and $\delta > 0$ such that  
\begin{equation*}
|V(\widehat{\pmb{\pi}}_n) - V(\pmb{\pi}^{\mathrm{opt}})| \le 
Kn^{-\delta}|\widehat{V}_{n} - V(\pmb{\pi}^{\mathrm{opt}})| + o_p(1/\sqrt{n}).
\end{equation*}
\end{lem}
\begin{cor}
Assume (C1)-(C3) and (AN1)-(AN8).  Then setting 
\begin{equation*}
n = \left\lceil \left\lbrace
\frac{
\Phi^{-1}(1-\zeta)\sigma^*
}{
\epsilon
}
\right\rbrace^2\right\rceil,
\end{equation*}
satisfies (OPT).  
\end{cor}
\noindent

\begin{rmrk}\label{markyMark}
  Given pilot or historical data, one can construct a plug-in
  estimator of $\sigma^{*2}$.     In the absence of such data, one 
  can use an elicited value for the variance of $Y$ under standard
  care as an {\em ad hoc} surrogate for $\sigma^{*2}$.    Heuristic
  justification for this surrogate is as follows.  If the variance of the outcome
  is at least as large under standard care as it is
  under  the optimal regime {\em and} the parametric
  estimator $\widehat{V}_{n}$ is at least as efficient as
  the sample mean of $n$ observations collected under the optimal
  policy, then 
  \begin{eqnarray*}
    \sigma^{*2} &=& \lim_{n\rightarrow\infty}\mathrm{Var}\left[
                    \sqrt{n}\left\lbrace \widehat{V}_{n} -
                    V(\pmb{\pi}^{\mathrm{opt}})
                    \right\rbrace
                    \right] \\
    & \le & \lim_{n\rightarrow \infty}\mathrm{Var}\left[
            \frac{1}{\sqrt{n}}\sum_{i=1}^{n}
            \left\lbrace 
            Y_i^*(\pmb{\pi}^{\mathrm{opt}}) -
            V(\pmb{\pi}^{\mathrm{opt}})
            \right\rbrace
            \right] \\
    &\le&  
           \lim_{n\rightarrow \infty}\mathrm{Var}\left[
            \frac{1}{\sqrt{n}}\sum_{i=1}^{n}
          \left\lbrace 
          Y_i^*(\pmb{\pi}^{\mathrm{soc}})
          -V(\pmb{\pi}^{\mathrm{soc}})
            \right\rbrace
          \right]
    \\ 
    &=& \mathrm{Var}\left\lbrace 
        Y^*(\pmb{\pi}^{\mathrm{soc}})
        \right\rbrace,
  \end{eqnarray*}
 where $\pmb{\pi}^{\mathrm{soc}}$ denotes standard of care.  If one
 is unwilling to make the above assumptions, an alternative would be
 to inflate the elicited value for $\mathrm{Var}\left\lbrace
   Y^*(\pmb{\pi}^{\mathrm{soc}})
   \right\rbrace$ by a constant factor.  
\end{rmrk}

\subsection{Projection-based sample size procedure}
Despite a deluge of new estimators of optimal treatment regimes
\citep[see][and references therein]{listBasedRegimes,
  zhou2017residual, laber2017functional, tao2017adaptive}, 
$Q$-learning with linear models remains
among the most commonly used methods in practice.  This popularity can be
partly attributed to: (i) the heavy use of linear models in seminal
papers on estimation of optimal treatment regimes
\citep[][]{murphyZThree, robinsTF, murphyZFive, mintron};
(ii) minimal requirements on the joint distribution of the
data-generating model; (iii) theoretical tractability
\citep[][]{chakraborty2009inference, moodieT,
  chakraborty2014inference, laber2014statistical}; and (iv) good
empirical performance even under some forms of misspecification
\citep[][]{schulte}.  Thus, our second sample size procedure is
designed for the setting where analysts plan to estimate the optimal
regime using $Q$-learning with linear models.  We do not assume that
the analysis model is correctly specified nor do we impose any
parametric structure on the generative model.  However, unlike the
procedure described in the preceding section, the resultant sample
size procedure derived here relies on quantities that would be
difficult to elicit from domain experts, we therefore require that one
has suitable pilot data available; such data could be historical or
collected as an internal pilot.

We assume that for $t=1,2$ one postulates models of the form
$Q_{t}(\bh_t, a_t;\mu_t) = \bh_{t,0}^{\T}\mu_{t,0} +
a_t\bh_{t,1}^{\T}\mu_{t,1}$,
where $\bh_{t,0},\bh_{t,1}$ are summaries of $\bh_{t}$ and
$\mu_t = (\mu_{t,0}^{\T},\mu_{t,1}^{\T})^{\T}$ are unknown parameters.
Define
$\mu_{2}^* = \arg\min_{\mu_2}P\left\lbrace Y-Q_{2}(\bH_2, A_2;\mu_2)
\right\rbrace^2$
and
$\mu_{1}^* = \arg\min_{\mu_1}P\big\lbrace \max_{a_2}Q_{2}(\bH_2,
a_2;\mu_{2}^*) \allowbreak - \allowbreak Q_{1}(\bH_1, A_1;\mu_1)
\big\rbrace^2$.
If (AN1) holds then, provided the requisite expectations exist,
$Q_2(\bh_2, a_2) = Q_2(\bh_2, a_2;\mu_2^*)$; however, even if
(AN1)-(AN4) hold, it need not follow that
$Q_1(\bh_1, a_1) = Q_1(\bh_1, a_1;\mu_1^*)$
\citep[][]{iqLearnSlagathor}.  
Nevertheless, it is still meaningful to discuss the 
mean outcome under the estimated optimal regime
when these models are misspecified.  
Define the optimal regime under linear
$Q$-learning with the above class of models 
as $\pmb{\pi}^{Q,\mathrm{opt}} = (\pi_1^{Q,\mathrm{opt}},
\pi_2^{Q,\mathrm{opt}})$ so that $\pi_{t}^{Q,\mathrm{opt}}(\bh_t) =
\arg\max_{a_t}
Q_{t}(\bh_t, a_t;\mu_t^*)$.  
\footnote{ 
The notation $\pmb{\pi}^{Q,\mathrm{opt}}$ is a bit
misleading in that
if the $Q$-functions are misspecified (which
we allow) it need not follow that 
 $V(\pmb{\pi}^{Q,\mathrm{opt}}) \ge V^{Q}(\mu_1, \mu_2)$
for all $(\mu_1,\mu_2)\in \Theta$; we do not assume 
that $\pmb{\pi}^{Q,\mathrm{opt}}$ satisfies such an inequality.   
 For additional
discussion, see \citet{mintron} and references therein.  
}
Define $\widehat{\mu}_{2,n} =
\arg\min_{\mu_2}
\pn\left\lbrace Y-Q_2(\bH_2, A_2;\mu_2)\right\rbrace^2$ 
and $\widehat{\mu}_{1,n}=\arg\min_{\mu_1}\big\lbrace
\max_{a_2}Q_2(\bH_2, a_2;\widehat{\mu}_{2,n}) - 
Q_{2}(\bH_{1}, A_1;\mu_1)
\big\rbrace^2$.  The estimated optimal decision
rule at time $t$ is $\widehat{\pi}_{t,n}^{Q}(\bh_t) =
\arg\max_{a_t} Q_{t}(\bh_t, a_t;\widehat{\mu}_{t,n})$.  

It is well-known that $V(\pmb{\pi}^{\mathrm{opt}})$ is not a smooth
functional of the generative model and consequently standard
approaches for inference, e.g., the bootstrap or series
approximations, will not hold without modification
\citep[][]{robinsTF, moodieT, chakraborty2009inference, mofn,
  chakraborty2014inference, laber2014statistical,
  luedtke2016statistical}.  To derive a test which satisfies (POW) we
invert a variant of a projection confidence interval
\citep[][]{berger1994p, robinsTF} for $V(\pmb{\pi}^{Q,\mathrm{opt}})$;
the interval we propose holds regardless of misspecification of the
$Q$-functions and does not require strong parametric assumptions on
the underlying generative model.   This approach requires
a confidence set for $(\mu_1^*, \mu_2^*)$ which we construct 
as follows.  Let $\bC_2$ be as in (AN7) and let 
$\widehat{\mathfrak{W}}_{2,n} = \left\lbrace \pn \bC_2
  \bC_2^{\T}\right\rbrace^{-1}\pn
\bC_2\bC_{2}^{\T}\left(Y-\bC_2^{\T}\widehat{\mu}_{2,n}\right)
\left\lbrace \pn \bC_2\bC_2^{\T}\right\rbrace^{-1}$ so
that $\mathfrak{Z}_{2,n,\varepsilon} = \Big\lbrace 
\mu_2\,:\, n(\mu_2 -
\widehat{\mu}_{2,n})^{\T}\widehat{\mathfrak{W}}_{2,n}^{-1}
(\mu_2 - \widehat{\mu}_{2,n}) \le \chi_{1-\varepsilon, \mathrm{dim}(\bC_2)}
\Big\rbrace$ is a Wald-type $(1-\varepsilon)\times 100$ confidence
set for $\mu_2^*$, where $\chi_{q,k}^2$ is the $q$th quantile of a
chi-square random variable with $k$ degrees of freedom.  For each $\mu_2$ define 
\begin{equation*}
\mu_{1}^*(\mu_2) = \arg\min_{\mu_1}\mathbb{E}\left\lbrace
\max_{a_2}Q_{2}(\bH_2, a_2;\mu_2) - Q_1(\bH_1, A_1;\mu_1)
\right\rbrace^2,
\end{equation*}
so that $\mu_{1}^*(\mu_2)$ is denotes the population-level for
the first-stage $Q$-function were it known that $\mu_2^* = \mu_2$;
thus, $\mu_1^* = \mu_1^*(\mu_2^*)$.   
Define $\widehat{\mu}_{1,n}(\mu_2) = 
\arg\min_{\mu_1}\pn\big\lbrace
\max_{a_2}Q_{2}(\bH_2, a_2;\mu_2) - Q_1(\bH_1, A_1;\mu_1)
\big\rbrace^2$, to be the least-squares estimator of
$\mu_1^*(\mu_2)$.  
Let $\bC_1 = (\bH_{1,0}^{\T}, A_1\bH_{1,1}^{\T})^{\T}$ and
define 
\begin{equation*}
\mathfrak{W}_{1,n}(\mu_2) = \left\lbrace \pn \bC_1 \bC_1^{\T}
\right\rbrace^{-1}\pn \bC_1\bC_1^{\T}\left\lbrace
\max_{a_2}Q_{2}(\bH_2, a_2;\mu_2) - \bC_1^{\T}\widehat{\mu}_{1,n}(\mu_2)
\right\rbrace 
\left\lbrace
\pn
\bC_1\bC_1^\T
\right\rbrace^{-1}.
\end{equation*}
A $(1-\varepsilon)\times 100\%$ Wald-type confidence set for
$\mu_1^*(\mu_2)$ is 
\begin{equation*}
\mathfrak{Z}_{1,n,\varepsilon}(\mu_2) = \left\lbrace
\mu_1\,:\, n\left[\mu_1-\widehat{\mu}_{1,n}(\mu_2)\right]^{\T}
\widehat{\mathfrak{W}}_{1,n}^{-1}(\mu_2)
\left[
\mu_1-\widehat{\mu}_{1,n}(\mu_2) 
\right] \le  \chi_{1-\varepsilon, \mathrm{dim}(\bC_1)}
\right\rbrace.
\end{equation*}
Thus, given $\varepsilon_1, \varepsilon_2 \in (0,1)$  with
$\vartheta = \varepsilon_1 + \varepsilon_2 \le 1$, a $(1-\vartheta)\times
100\%$ confidence set for $(\mu_1^*, \mu_2^*)$ is 
$\Xi_{n, 1-\vartheta} = 
\left\lbrace
(\mu_1,\mu_2) \,:\, \mu_2 \in
  \mathfrak{Z}_{2,n,1-\varepsilon_2}\,\, \mbox{ and } \,\, 
\mu_1 \in\mathfrak{Z}_{1,n,1-\varepsilon_1}(\mu_2)
\right\rbrace.$

For each $\mu_1, \mu_2$ define
\begin{multline*}
\delta(\mu_1,\mu_2)  = 
4Y1_{A_1\bH_{1,1}^{\T}\mu_{1,1} > 0}1_{A_2\bH_{2,1}^{\T}\mu_{2,1} > 0}
+ 2\left\lbrace
1_{A_{1}\bH_{1,1}^{\T}\mu_{1,1} \le 0} - (1/2)
\right\rbrace 
\max_{a_1}Q_1(\bH_1, a_1;\mu_1)  \\
+ 41_{A_{1}\bH_{1,1}^{\T}\mu_{1,1} > 0}\left\lbrace
1_{A_{2}\bH_{2,1}^{\T}\mu_{2,1} \le 0} - (1/2)
\right\rbrace \max_{a_2}Q_{2}(\bH_{2}, a_2;\mu_{2}),
\end{multline*}
and subsequently define
$\widehat{V}_{n}^{Q}(\mu_1, \mu_2) = \pn \delta(\mu_1, \mu_2)$
and its population-level analog $V^{Q}(\mu_1, \mu_2) = 
\mathbb{E}\delta(\mu_1, \mu_2)$.  
Then, $\widehat{V}_{n}(\widehat{\mu}_{n,1},\widehat{\mu}_{2,n})$ is
the augmented inverse probability weighted estimator
of $V(\pmb{\pi}^{Q,\mathrm{opt}})$ \citep[][]{baqun2} and
$V^Q(\mu_1^*, \mu_2^*) = V(\pmb{\pi}^{Q,\mathrm{opt}})$ 
\cite[see also][]{mintron, yingqi2}.  
Define
$\varsigma^2(\mu_1,\mu_2) = 
\mathbb{E}
\left\lbrace
\delta(\mu_1,\mu_2) - \mathbb{E}\delta(\mu_1,\mu_2)
\right\rbrace^2$ and 
$\widehat{\varsigma}_{n}^2 = 
\mathbb{P}_n
\left\lbrace
\delta(\mu_1,\mu_2) - \mathbb{P}_n\delta(\mu_1,\mu_2)
\right\rbrace^2$.
For
any fixed $(\mu_1, \mu_2)$, 
it follows that
\begin{equation*}
\sqrt{n}\left\lbrace
\widehat{V}_{n}^{Q}(\mu_1,\mu_2) - V^{Q}(\mu_1, \mu_2)
\right\rbrace \leadsto \mathrm{Normal}\left\lbrace 
0, \varsigma^2(\mu_1,\mu_2)
\right\rbrace,
\end{equation*}
provided that $\mathbb{E}\delta^2(\mu_1,\mu_2) < \infty$.  
Choose $\vartheta_1$ and $\vartheta_2$ such that $\vartheta_1
+\vartheta_2 = \alpha$, 
then the proposed
$\alpha$-level test for (POW) rejects when 
\begin{equation*}
\inf_{(\mu_1,\mu_2)\in \Xi_{n,1-\vartheta_1}}\left[
\widehat{V}_{n}^{Q}(\mu_1,\mu_2) - \frac{
z_{1-\vartheta_2}\widehat{\varsigma}_{n}(\mu_1,\mu_2)}
{
\sqrt{n}
}
\right] \ge B_0.
\end{equation*}
Under the null, 
$V(\pmb{\pi}^{Q,\mathrm{opt}}) \le B_0$, so that the type 
I error is bounded above by 
\begin{multline*}
P\left\lbrace
\inf_{(\mu_1,\mu_2)\in \Xi_{n,1-\vartheta_1}}\left[
\widehat{V}_{n}^{Q}(\mu_1,\mu_2) - \frac{
z_{1-\vartheta_2}\widehat{\varsigma}_{n}(\mu_1,\mu_2)
}{
\sqrt{n}
}
\right]
\ge V(\pmb{\pi}^{Q,\mathrm{opt}})
\right\rbrace \smallskip \\ 
\le 
P\left\lbrace
\widehat{V}_{n}^Q(\mu_1^*, \mu_2^*) 
- \frac{
z_{1-\vartheta_2}\widehat{\varsigma}_{n}(\mu_1^*,\mu_2^*)
}{
\sqrt{n}
}
\ge V(\pmb{\pi}^{Q,\mathrm{opt}})
\right\rbrace + \vartheta_1 + o(1) \smallskip \\
= P\left\lbrace
\sqrt{n}\left[
\frac{
\widehat{V}_{n}^{Q}(\mu_1^*, \mu_2^*) - V^Q(\mu_1^*,\mu_2^*)
}{
\widehat{\varsigma}_{n}(\mu_1^*,\mu_2^*)
}
\right]
\ge z_{1-\vartheta_2}
\right\rbrace + \vartheta_1+o(1) 
\smallskip \\ 
\le \vartheta_1 + \vartheta_2 + o(1),
\quad\quad\quad\quad\quad\quad\quad\quad
\quad\quad\quad\quad\quad\quad\quad\quad
\quad\quad\quad\quad\quad\quad\quad\quad
\end{multline*}
where the first inequality follows from 
$P\left\lbrace (\mu_1^*,\mu_2^*) \in
  \Xi_{n,1-\vartheta_1}\right\rbrace  \ge 1-\vartheta_1 + o(1)$.  
If 
$\widehat{\varsigma}_{n}(\mu_1,\mu_2) > 0$ with probability one for all 
$(\mu_1,\mu_2) \in \Theta$, then it can be seen that 
the power of the proposed test is  
\begin{align}
& P\left\lbrace
\inf_{(\mu_1,\mu_2)\in \Xi_{n,1-\vartheta_1}}\left[
\widehat{V}_{n}^{Q}(\mu_1,\mu_2) - \frac{
z_{1-\vartheta_2}
\widehat{\varsigma}_{n}(\mu_1,\mu_2)
}{
\sqrt{n}
}
\right]
\ge B_0
\right\rbrace \quad\quad \smallskip \notag \\
\begin{split}
& \quad\quad = 
P\Bigg\lbrace 
\inf_{(\mu_1,\mu_2)\in \Xi_{n,1-\vartheta_1}}\Bigg[
\frac{
\sqrt{n}\left\lbrace
\widehat{V}_{n}^{Q}(\mu_1,\mu_2) - V^{Q}(\mu_1,\mu_2)
\right\rbrace
}{
\widehat{\varsigma}_{n}(\mu_1,\mu_2)
}\\
&\quad\quad \quad\quad  \quad\quad  \quad\quad + 
\frac{
\sqrt{n}\left\lbrace
V^Q(\mu_1,\mu_2) -B_0
\right\rbrace
}{
\widehat{\varsigma}_{n}(\mu_1,\mu_2)
}
\Bigg]
 \ge z_{1-\vartheta_2}
\Bigg\rbrace   \label{firstBound}  
\end{split} \\
\begin{split}
& \quad\quad \ge 
P\Bigg\lbrace 
\inf_{(\mu_1,\mu_2)\in \Xi_{n,1-\vartheta_1}}\Bigg(
\frac{
\sqrt{n}\left\lbrace
\widehat{V}_{n}^{Q}(\mu_1,\mu_2) - V^{Q}(\mu_1,\mu_2)
\right\rbrace
}{
\widehat{\varsigma}_{n}(\mu_1,\mu_2)
}\\
&\quad\quad  \quad\quad  \quad\quad  \quad\quad + 
\frac{
\min\left[
\sqrt{n}\left\lbrace
V^Q(\mu_1,\mu_2) -B_0
\right\rbrace,\,\sqrt{n}\eta
\right]
}{
\widehat{\varsigma}_{n}(\mu_1,\mu_2)
}
\Bigg)
 \ge z_{1-\vartheta_2}
\Bigg\rbrace. \label{secondBound}
\end{split}
\end{align}
The minimum in (\ref{secondBound}) is analogous to plugging-in the 
smallest possible difference under the alternative 
$V(\pmb{\pi}^{Q,\mathrm{opt}}) -B_0 \ge \eta$; see
Remark \ref{markyMark} for additional discussion.  
The sampling distribution of the test statistic under the
alternative is complex and difficult to approximate using
series approximations; thus, to estimate a sample size that
will yield the desired power, we use the bootstrap.

\subsubsection{Bootstrap power calculation}
We assume that one has available pilot data
$\mathcal{D}_{n_0} = \left\lbrace (\bX_{1,i}, A_{1,i}, \bX_{2,i},
  A_{2,i}, Y_i)\right\rbrace_{i=1}^{n_0}$
comprising $n_0$ $i.i.d.$ trajectories from the same population from
which trial participants will be drawn. We estimate the power required
in (POW) using the bootstrap with a resample size of $n \ge n_0$ and
solve for the smallest $n$ such that the estimated power exceeds a 
given threshold.   
In our asymptotic analyses, we let both $n$ and $n_0$ diverge to infinity;
however, as we anticipate the trial sample size to be much larger than
that of the  pilot, we focus on an asymptotics in which $n$ goes
to infinity ``first.''   We assume:
\begin{itemize}
\item[(PR1)] $\mathbb{E}Y^2||\bC_2||^2 < \infty$ and
  $\mathbb{E}||\bC_1||^2||\bC_2||^2 < \infty$;
\item[(PR2)] $\mathbb{E}\bC_1\bC_1^{\T}$ and
  $\mathbb{E}\bC_2\bC_2^{\T}$
are finite and strictly positive definite;
\item[(PR3)]  $\inf_{(\mu_1,\mu_2)\in\Theta}\mathbb{E}
\left\lbrace
\delta(\mu_1, \mu_2) - \mathbb{E}\delta(\mu_1, \mu_2)
\right\rbrace^2 > 0$ and 
$\sup_{(\mu_1,\mu_2)\in\Theta}\mathbb{E}
\left\lbrace
\delta(\mu_1, \mu_2) - \mathbb{E}\delta(\mu_1, \mu_2)
\right\rbrace^2 < \infty$;
\item[(PR4)]  the classes $\mathcal{F}_1 = 
 \left\lbrace \delta(\mu_1,\mu_2)\,:\, (\mu_1,\mu_2)\in\Theta
 \right\rbrace$ and $\mathcal{F}_2 = \left\lbrace
 \delta^2(\mu_1,\mu_2)\,:\, (\mu_1,\mu_2)\in\Theta
 \right\rbrace$ are Donsker;
\item[(PR5)] $\mathbb{E}\delta(\mu_1,\mu_2)$ is uniformly continuous
in a neighborhood of $(\mu_1^*, \mu_2^*)$.    
\end{itemize}
The foregoing assumptions are standard in linear $Q$-learning 
and mirror those used in linear regression
\citep[][]{laber2014statistical}.


Let $\mathbb{P}_{n,n_0}^{(b)}$ denote the bootstrap empirical
distribution corresponding to a resample size of $n$.  For any
functional $Z_n = f(P, \mathbb{P}_{n_0})$ we define its bootstrap analog
$Z_{n_0,n}^{(b)} = f\left\lbrace \mathbb{P}_{n_0}, \mathbb{P}_{n,n_0}^{(b)}\right\rbrace$.
Let $P_B$ denote probabilities computed with respect the 
bootstrap distribution conditional on the pilot data.
The bootstrap estimator of the sample size required for
(POW) is the positive integer $n$ which solves 
\begin{multline*}
P_B\left\lbrace\rule{0cm}{1.15cm}\right.
\inf_{(\mu_1,\mu_2)\in \Xi_{n_0, n, 1-\vartheta_1}^{(b)}}
\left(\rule{0cm}{1.05cm}\right.
\frac{
\sqrt{n}\left\lbrace
\widehat{V}_{n_0,n}^{Q(b)}(\mu_1, \mu_2) - \widehat{V}_{n_0}^{Q}(\mu_1, \mu_2)
\right\rbrace 
}{
\widehat{\varsigma}_{n_0,n}^{(b)}(\mu_1,\mu_2)
}\\
+ 
\frac{
\min\left[
\sqrt{n}\left\lbrace
\widehat{V}_{n_0}^{Q}(\mu_1,\mu_2) - B_0
\right\rbrace,
\,\sqrt{n}\eta\right]
}{
\widehat{\varsigma}_{n_0,n}^{(b)}(\mu_1,\mu_2)
}
\left. \rule{0cm}{0.95cm}\right) 
\ge z_{1-\vartheta_2} 
\left. \rule{0cm}{1.05cm}\right\rbrace \ge 1-\gamma 
\end{multline*}
where $\vartheta_1 + \vartheta_2 = \alpha$; the probability on the left hand side
of the inequality 
can be computed to desired precision by Monte Carlo methods.  
The following results establish consistency of the bootstrap as 
$n_0$ and $n$ diverge. 
\begin{thm}
Assume (C1)-(C3) and (PR1)-(PR4).  Let $\vartheta_1 \in (0,1)$ be
fixed.  Let $\varkappa, K > 0$ be arbitrary, then
\begin{multline*}
\lim_{n,n_0\rightarrow \infty}
P\Bigg(
\sup_{|v| \le K}\Bigg|
P_B\left[
\inf_{(\mu_1,\mu_2)\in \Xi_{n_0, n, 1-\vartheta_1}^{(b)}} 
\frac{
\sqrt{n}\left\lbrace
\widehat{V}_{n_0,n}^{Q(b)}(\mu_1, \mu_2) - \widehat{V}_{n_0}^{Q}(\mu_1, \mu_2)
\right\rbrace 
}{
\widehat{\varsigma}_{n_0,n}^{(b)}(\mu_1,\mu_2)
}
\ge v
\right]\\ 
- 
P\left[
\inf_{(\mu_1,\mu_2)\in\Xi_{n,1-\vartheta_1}}
\frac{
\sqrt{n}\left\lbrace
\widehat{V}_{n}^{Q}(\mu_1, \mu_2) - V^{Q}(\mu_1, \mu_2)
\right\rbrace 
}{
\widehat{\varsigma}_{n}(\mu_1,\mu_2)
}
\ge v
\right]
\Bigg| > \varkappa
\Bigg) = 0.
\end{multline*}
\end{thm}
\noindent
The preceding result does not include 
$\sqrt{n}\left\lbrace V^Q(\mu_1, \mu_2)-B_0\right\rbrace/
\widehat{\varsigma}_{n}(\mu_1,\mu_2)$ 
(or its bootstrap analog) because, under the
alternative, provided
$V(\mu_1,\mu_2) > B_0$ for all $(\mu_1,\mu_2)$ in 
a sufficiently small neighborhood of $(\mu_1^*, \mu_2^*)$, 
this term (and its bootstrap analog) will diverge to infinity
so that the conclusion of the above theorem will hold
trivially.  The following result characterizes the limiting 
tail behavior of this term; the factor of $\sqrt{n}$ on the
right-hand-side
of each probability assignment reflects the fact that, under the
alternative, we expect $\sqrt{n}\left\lbrace
  V^{Q}(\pi^{Q,\mathrm{opt}}) - B_0
\right\rbrace$ to diverge at rate $\sqrt{n}$.  
\begin{thm}
Assume (C1)-(C3) and (PR1)-(PR4). Let
$\vartheta_1 \in (0,1)$ and $\eta \ge 0$ be fixed.  
In addition, assume that $n_0\rightarrow \infty$ as 
$n\rightarrow \infty$ and that there exists $c>0$ so that
$\inf_{(\mu_1,\mu_2)\in \Omega}\widehat{\varsigma}_{n}(\mu_1,\mu_2)
\ge c.$ 
 Let $\varkappa, K > 0$ be arbitrary then
\begin{multline*}
\lim_{n,n_0\rightarrow\infty}P\Bigg(
\sup_{|v| \le K}\Bigg| P_{B}\left[
\inf_{(\mu_1,\mu_2)\in\Xi_{n_0,n,1-\vartheta_1}^{(b)}}\frac{
\min\left[
\sqrt{n}\left\lbrace
\widehat{V}_{n_0,n}^{Q,(b)}(\mu_1,\mu_2) - B_0
\right\rbrace,
\,\sqrt{n}\eta 
\right]
}{
\widehat{\varsigma}_{n_0,n}^{(b)}(\mu_1,\mu_2)
}
\ge \sqrt{n}v 
\right]\\
- P\left[
\inf_{(\mu_1,\mu_2)\in\Xi_{n,1-\vartheta_1}}
\frac{
\min\left[
\sqrt{n}\left\lbrace
\widehat{V}_{n}^{Q}(\mu_1,\mu_2)-B_0
\right\rbrace,
\,\sqrt{n}\eta
\right]
}{
\widehat{\varsigma}_{n}(\mu_1,\mu_2)
}
\ge \sqrt{n}v
\right]
\Bigg|
> \varkappa
\Bigg) = 0.
\end{multline*}
\end{thm}

To choose $n$ so that (OPT) holds asymptotically we make use of the
following bound.  For any (possibly data-dependent) 
sequence $(\widetilde{\mu}_{1,n}, \widetilde{\mu}_{2,n}) \in
\Xi_{n, 1-\vartheta_1}$ such that 
$\widehat{V}_{n}^Q(\mu_1^*, \mu_2^*) \le 
\widehat{V}_{n}^{Q}(\widetilde{\mu}_{1,n}, \widetilde{\mu}_{2,n}) +
o_P(1/\sqrt{n})$ it follows that  
\begin{multline*}
P\Bigg[
 V^Q(\widetilde{\mu}_{1,n}, \widetilde{\mu}_{2,n}) \ge 
 V^Q(\pmb{\pi}^{Q,\mathrm{opt}}) + 
 \inf_{(\mu_1, \mu_2) \in \Xi_{n, 1-\vartheta_1}}\left\lbrace
 \widehat{V}_{n}^{Q}(\mu_1, \mu_2) - V^Q(\mu_1, \mu_2)
 \right\rbrace \\ 
 - \sup_{(\mu_1,\mu_2)\in \Xi_{n,1-\vartheta_1}}\left\lbrace
 \widehat{V}_{n}^{Q}(\mu_1, \mu_2) - V^Q(\mu_1, \mu_2)
 \right\rbrace
\Bigg] \ge 1-\vartheta_1 + o(1).
\end{multline*}
Thus, if $\mathfrak{Q}_{n, 1-\vartheta_2, 1-\vartheta_1}$ is the $(1-\vartheta_2)$
quantile  of 
\begin{equation*}
\inf_{(\mu_1, \mu_2) \in \Xi_{n, 1-\vartheta_1}}\left\lbrace
 \widehat{V}_{n}^{Q}(\mu_1, \mu_2) - V^Q(\mu_1, \mu_2)
 \right\rbrace  
 - \sup_{(\mu_1,\mu_2)\in \Xi_{n,1-\vartheta_1}}\left\lbrace
 \widehat{V}_{n}^{Q}(\mu_1, \mu_2) - V^Q(\mu_1, \mu_2)
 \right\rbrace
\end{equation*} 
then choosing $\vartheta_1 + \vartheta_2 \le \zeta$ and $n$ such that
$\mathfrak{Q}_{n,1-\vartheta_1, 1-\vartheta_2}/\sqrt{n} \le \epsilon$ ensures that 
(OPT) holds asymptotically.   Of course,
$\mathfrak{Q}_{n,, 1-\vartheta_1, 1-\vartheta_2}$ is unknown so we estimtate it using
the bootstrap, i.e., we select $n$ so that 
$\mathfrak{Q}_{n_0, n, 1-\vartheta_1, 1-\vartheta_2}^{(b)}/\sqrt{n}
\le \epsilon$.  

\begin{rmrk}\label{markyMark}
To estimate the power at a given sample size one could use the
bootstrap analog of  (\ref{firstBound}).  Indeed, the preceding
theoretical results can be easily modified to hold without the
$\min$ operation.  However, the required sample size derived from 
(\ref{firstBound}) will be based on an {\em estimated effect size} rather
than the minimal effect size of interest, $\eta$. A consequence 
of using the estimated effect size is that  
as the true effect size increases the estimated required sample size
will decrease keeping the power fixed at (approximately)
$(1-\gamma)\times 100$. 
However, in application, it desirable to have power $(1-\gamma)\times
100$ at effect size $\eta$ but larger power if the effect size exceeds
$\eta$.  Taking the minimum, as in (\ref{secondBound}), ensures that
the power diverges to one as the true effect size grows large. 
\end{rmrk}

\section{Simulation experiments}
We examine the finite sample performance of the proposed sample size
procedures using a series of simulation experiments.  Performance is
measured in terms of the proposed criteria (POW) and (OPT). For each
generative model, we also compute the number of samples required to
compare the mean outcomes under standard care to that under the fixed
regimes $\pmb{\pi}^{i,j},\,i,j\in \lbrace-1,1\rbrace$, where
$\pi_1^{i,j}(\bh_1) \equiv i$ and
$\pi_2^{i,j}(\bh_2) \equiv j$. This comparison allows us to
evaluate how much the sample size must be inflated to estimate and/or
evaluate an optimal dynamic treatment regime relative to the comparison 
of fixed and embedded regimes \citep[][]{almirall2012designing}.  

We first consider a generative model in which the assumptions
(AN1)-(AN8) 
for the normality-based sample size procedure hold.  
This generative model is as follows:
\begin{equation*}
\begin{array}{ll}
\bX_1 \sim N_{4}\{0, \Omega_{AR1}(0.5)\}, 
& \bH_{1,0}^T = (1, X_{1,0}),\\
\bH_{1,1}^T = (1, X_{1,1}), 
&\bH_{1,2}^T = (1,X_{1,2}), \\
 \bH_{1,3}^T = (1, X_{1,3}), 
& A_1,A_2 \sim_{i.i.d.} \mathrm{Unif}\{-1, 1\},\\ 
{\phi}_1,\phi_2, \upsilon \sim_{i.i.d.} N(0,1), & 
X_{2,0} = \bH_{1,0}^T{\mu}_{1,0}^* + A_1\bH_{1,1}^T\mu_{1,1}^* + {\phi}_1,\\ 
X_{2,1} = \bH_{1,2}^T\mu_{2,0}^* + A_1\bH_{1,3}^T\mu_{2,1}^* + {\phi}_2,  &
\bH_{2,0}^T = (1, X_{1,0}, A_1, X_{2,0}), \\  
\bH_{2,1}^T = 
(1, X_{1,2}, A_1, X_{2,1}) ,
 &
Y = \bH_{2,0}^T\beta_{2,0}^* + A_2\bH_{2,1}^T\beta_{2,1}^* + \upsilon, \\
\end{array}
\end{equation*}
where $\Omega_{AR1}(0.5)$ is an autoregressive covariance matrix such
that $\{\Omega_{AR1}(0.5)\}_{ij} = 0.5^{|i-j|}$. 

Let $\pmb{\pi}^{\mathrm{fixed,opt}}$ denote the optimal fixed regime
such that $\pmb{\pi}^{\mathrm{fixed,opt}} = \pmb{\pi}^{i^*,j^*}$, where 
$i^*,j^* = \arg\max_{i,j\in\lbrace-1,1\rbrace} 
V(\pmb{\pi}^{i,j})$. We examine the performance of the proposed
methods under
parameter
values which result in the following relationships between
$\pmb{\pi}^{\mathrm{fixed,opt}}$, $\pmb{\pi}^{\mathrm{opt}}$, and
$B_0$:
\begin{enumerate}
\item $V(\pmb{\pi}^{\mathrm{opt}}) = V(\pmb{\pi}^{\mathrm{fixed,opt}})
  = B_0 +
 \eta$;
\item $V(\pmb{\pi}^{\mathrm{opt}}) - 0.5\eta = V(\pmb{\pi}^{\mathrm{fixed,opt}}) = B_0 + \eta$;
\item $V(\pmb{\pi}^{\mathrm{opt}}) - \eta = 
V(\pmb{\pi}^{\mathrm{fixed,opt}}) = B_0 + \eta$;
\item $V(\pmb{\pi}^{\mathrm{opt}}) - 2\eta = 
V(\pmb{\pi}^{\mathrm{fixed,opt}}) = B_0 + \eta$.
\end{enumerate}
Define
$\Delta = \left\lbrace V(\pmb{\pi}^{\mathrm{opt}}) -
  V(\pmb{\pi}^{\mathrm{fixed,opt}})\right\rbrace /\eta$
to be a measure of benefit associated with implementing an optimal
dynamic treatment regime relative to the optimal fixed regime.  It can
be seen that $\Delta$ ranges from zero to two across the above
scenarios. Parameter settings indexing the generative model which
yield these values of $\Delta$ when $\eta=1$ are:
\begin{enumerate}
\item $\mu_{1,0} = (-1, 1)$, $\mu_{1,1} = (4, 1)$, 
$\mu_{2,0} = (-0.4, -1)$, $\mu_{2,1} = (4, 1)$, $\beta_{2,0}^* = (0.5, 0.5, -1 ,1)$ and $\beta_{2,1}^* = (1, 0.5, 0.5, 1)$;
\item $\mu_{1,0} = (-1, 1)$, $\mu_{1,1} = (4, 1)$, 
$\mu_{2,0} = (-0.4, -1)$, $\mu_{2,1} = (-4, 1)$, $\beta_{2,0}^* = (0.5, 0.5, -1 ,1)$ and $\beta_{2,1}^* = (1, -0.9, 0.5, 1)$;
\item $\mu_{1,0} = (-1, 1)$, $\mu_{1,1} = (4, 1)$, $\mu_{2,0} = (-0.4, -1)$, $\mu_{2,1} = (-4, 1)$, $\beta_{2,0}^* = (0.5, 0.5, -1 ,1)$ and $\beta_{2,1}^* = (1, -1.75, 0.5, 1)$;
\item $\mu_{1,0} = (-1, 1)$, $\mu_{1,1} = (4, 1)$, $\mu_{2,0} = (-0.4, -1)$, $\mu_{2,1} = (-4, 1)$, $\beta_{2,0}^* = (0.5, 0.5, -1 ,1)$ and $\beta_{2,1}^* = (1, -3.25, 0.5, 1)$.
\end{enumerate}

Let $\gamma = 0.1$, $\alpha = 0.05$, $\zeta = 0.1$, and
$\epsilon = 0.3$. Thus, if (POW) holds, then an $\alpha$-level test of
$H_0: V(\pmb{\pi}^{\mathrm{opt}}) \le B_0$ will have approximately
90\% power, and if (OPT) holds, then
$P\left[ \mathbb{E}\left\lbrace
    Y^*(\widehat{\pmb{\pi}}_{n})\big|\mathcal{D}_{n} \right\rbrace \ge
  V(\pmb{\pi}^{\mathrm{opt}}) - 0.3 \right] \ge 0.9 + o(1)$.
Recall that the normality-based sample size procedure requires
specification of $\sigma^*$. We consider three possibilities: (i)
$\sigma^*$ is known, e.g., correctly elicited from domain experts;
(ii) $\sigma^*$ is estimated using a pilot study of $n_0$ patients;
and (iii) $\sigma^*$ is estimated using the {\em ad hoc} procedure
presented in Remark \ref{markyMark} using a sample of $n_0$ subjects
treated under standard care wherein we assume that patients are
assigned the optimal treatment 80\% of the time and suboptimal
treatment the remaining 20\% of the time.  To form a baseline
for comparison, we also compute the sample size required to power
a test of the null ${V}(\pmb{\pi}^{\mathrm{fixed,opt}}) \le
B_0$
against the alternative $V(\pmb{\pi}^{\mathrm{fixed,opt}}) > B_0$
where
it is assumed that $\pmb{\pi}^{\mathrm{fixed, opt}}$ is 
known {\em a priori} as is 
$\mathrm{Var}\left\lbrace 
Y^*(\pmb{\pi}^{\mathrm{fixed,opt}}) \right\rbrace$;  
this reflects the common practice of comparing
a fixed regime against standard of care or another fixed regime.
All results are based on
500 Monte Carlo replications.  

Table \ref{normTablePOW} displays the
average estimated sample size and its operating characteristics 
across the four settings of the proposed
generative model and three approaches to selecting $\sigma^*$ when sizing for just condition (POW).
Table \ref{normTableOPT} displays the same results when sizing for condition (OPT). A table of results 
when sizing for both jointly is contained in the Supplemental Materials.
In the case where $\Delta = 0$, the optimal treatment 
regime provides no benefit over the optimal fixed 
regime, thus, this setting reflects a worst-case in terms of
the conservatism of sizing for (POW) and (OPT) rather than 
simply sizing to identify the optimal fixed regime;
in this case, the proposed sample size procedure attains
nominal levels for (POW) and (OPT) at the cost of an inflated sample
size.  However, as $\Delta$ increases, so that the benefits 
of personalizing treatment relative to a fixed regime also
increase,  it can be seen that the sample size required for (POW)
and (OPT) can (perhaps surprisingly) be considerably smaller than required for 
identifying an optimal embedded regime provided that one has a 
high-quality estimate for $\sigma^*$ either through elicitation or 
a pilot study.  Table \ref{normTableOPT} has power 1.0 for all
cases considered which is a consequence of using an upper
bound on the difference between the value of the estimated
regime and the optimal regime. One could potentially
explore data-adaptive adjustments, e.g., the double bootstrap,
to reduce this excess power.  

    
\begin{table}[ht]
	\caption{
		Estimated power (POW) under a correctly
		specified generative model using the normality-based sample
		size procedure at a nominal level of 90.  To form a baseline for
		comparison, $\widehat{n}^{\mathrm{fixed}}$, shows the
		required sample size to compare the optimal embedded regime
		with standard of care. }
	\begin{center}
		\begin{tabular}{| c c c c c c c c c |} \hline
			$\Delta$ & Method for $\sigma^*$ & $n_0$ & (POW) & (OPT) 
			& $\hat{n}^{\mathrm{fixed}}$ & $\mathbb{E}\hat{n}$ & $\mathrm{Med}(\hat{n})$ & $\mathrm{SD}\hat{n}$ \\ \hline
			\hline  
			0 & known & 50 & 90.0 & - & 74 & 130 & $-$ & $-$ \\
			0 & pilot study & 50 & 87.2 & - & 74 & 100.57 & 99 & 26.74  \\
			0 & surrogate & 50 & 98.6 & - & 74 & 186.44 & 185 & 44.65 \\ \hline
			0.5 & known & 50 & 100 & - & 111 & 124 & $-$ & $-$ \\
			0.5 & pilot study & 50 & 99.8 & - & 111 & 132.14 & 131 & 29.39 \\
			0.5 & surrogate & 50 & 100 & - & 111 & 164.06 & 159 & 46.60 \\ \hline
			1 & known & 50 & 100 & - & 151 & 134 & $-$ & $-$ \\
			1 & pilot study & 50 & 100 & - & 151 & 160.43 & 158 & 33.11 \\
			1 & surrogate & 50 & 100 & - & 151 & 203.78 & 199.5 & 62.16  \\
			\hline 
			2 & known & 50 & 100 & - & 251 & 165 & $-$ & $-$ \\
			2 & pilot study & 50 & 100 & - & 251 & 235.47 & 233 & 46.14 \\
			2 & surrogate & 50 & 100 & - & 251 & 280.04 & 275  & 82.16 \\
			\hline
		\end{tabular}
	\end{center}
	\label{normTablePOW}
\end{table}    

\begin{table}[ht]
	\caption{
		Estimated concentration (OPT) under a correctly
		specified generative model using the normality-based sample
		size procedure at a nominal level of 90.  To form a baseline for
		comparison, $\widehat{n}^{\mathrm{fixed}}$, shows the
		required sample size to compare the optimal embedded regime
		with standard of care. }
	\begin{center}
		\begin{tabular}{| c c c c c c c c c |} \hline
			$\Delta$ & Method for $\sigma^*$ & $n_0$ & (POW) & (OPT)
			& $\hat{n}^{\mathrm{fixed}}$ & $\mathbb{E}\hat{n}$ & $\mathrm{Med}(\hat{n})$ & $\mathrm{SD}\hat{n}$ \\ \hline
			\hline  
			0 & known & 50 & - & 100 & 74 & 277 & $-$ & $-$ \\
			0 & pilot study & 50 & - & 100 & 74 & 204.51 & 199.5 & 60.60  \\
			0 & surrogate & 50 & - & 100  & 74 & 394.63 & 390 & 90.04 \\ \hline
			0.5 & known & 50 & - & 100 & 111 & 263 & $-$ & $-$ \\
			0.5 & pilot study & 50 & - & 100  & 111 & 269.91 & 264 & 62.58 \\
			0.5 & surrogate & 50 & - & 100 & 111 & 353.35 & 343.5 & 98.09 \\ \hline
			1 & known & 50 & - & 100 & 151 & 285 & $-$ & $-$ \\
			1 & pilot study & 50 & - & 100 & 151 & 335.69 & 332.5 & 69.93 \\
			1 & surrogate & 50 & - & 100 & 151 & 431.93 & 420 & 130.70  \\
			\hline 
			2 & known & 50 & - & 100 & 251 & 352 & $-$ & $-$ \\
			2 & pilot study & 50 & - & 100 & 251 & 495.52 & 491 & 92.61 \\
			2 & surrogate & 50 & - & 100 & 251 & 605.82 & 578.5 & 176.10 \\
			\hline
		\end{tabular}
	\end{center}
	\label{normTableOPT}
\end{table}

We also examined the performance of the normality-based sample size
when the postulated modeling assumptions are violated. For these
simulations, we used the following generative model:
\begin{equation*}
\begin{array}{ll}
	X_1 \sim \mathrm{N}(0,1), & A_1, A_2\sim_{i.i.d.} 
\mathrm{Unif}\{-1, 1\},  \\
\phi \sim t_3,
	& X_2 = \mu_0^* + \mu_1^*X_1 + \mu_2^*A_1 + \mu_3^*A_1X_1 + \mu_4^*X_1^2 + \phi,\\
	\bH_{2,0} = (1, X_1, A_1, X_1 A_1, X_2), & 
\bH_{2,1} = (1, A_1, X_2), \\	
\upsilon \sim N(0,1),
& Y = \bH_{2,0}^T \beta_{2,0}^* + A_2\bH_{2,1}^T \beta_{2,1}^* + \upsilon.
\end{array}
\end{equation*}
As previously, we let $V(\pmb{\pi}^{\mathrm{opt}}) - \Delta\eta =
V(\pmb{\pi}^{\mathrm{fixed,opt}}) = B_0 + \eta$ and consider $\Delta
\in \{0, 0.5, 1, 2\}$. We set $\mu^* = (1, 0.5, 0.5, 0.1, 1)$ and choose 
$\beta^*$ as follows: 
\begin{enumerate}
	\item $\Delta = 0$, $\beta_{2,0}^* = (1, 0.5, 0.5, 0.5, 1.5)$, $\beta_{2,1}^* = (-1, -1, 0)$;
	\item $\Delta = 0.5$, $\beta_{2,0}^* = (1, 0.5, 0.5, 0.5, 1.5)$, $\beta_{2,1}^* = (-1, -1, 0.55)$;
	\item $\Delta = 1$, $\beta_{2,0}^* = (1, 0.5, 0.5, 1, 1.5)$, $\beta_{2,1}^* = (-1, -1, 0.65)$;
	\item $\Delta = 2$, $\beta_{2,0}^* = (1, 0.5, 0.5, 2.3, 1.5)$, $\beta_{2,1}^* = (-1, -1, 0.71)$.
\end{enumerate}

The average sample size and operating characteristics of the
normality-based sample size procedure when sizing for condition (POW) 
are displayed in  Table
\ref{quadTablePOW} whereas the average sample size and characteristics
for condition (OPT) are in 
Table \ref{quadTableOPT}. 
Results for sizing to guarantee  both 
conditions 
jointly are contained in the Supplemental Materials.
The proposed method continues to attain nominal
levels for $\Delta \ge 1$, but is underpowered when using a pilot study 
to estimate $\sigma^*$ and there is little or no benefit to the optimal 
regime over the optimal embedded regime.


\begin{table}[ht]
	\caption{
		Estimated power (POW) under a model which violated the normality assumptions using the normality-based sample size procedure at a nominal level of 90.  To form a baseline for
		comparison, $\widehat{n}^{\mathrm{fixed}}$, shows the
		required sample size to compare the optimal embedded regime
		with standard of care. }
	\begin{center}
		\begin{tabular}{| c c c c c c c c c |} \hline
			$\Delta$ & Method for $\sigma^*$ & $n_0$ & (POW) & (OPT) 
			& $\hat{n}^{\mathrm{fixed}}$ & $\mathbb{E}\hat{n}$ & $\mathrm{Med}(\hat{n})$ & $\mathrm{SD}\hat{n}$ \\ \hline
			\hline  
			0 & known & 50 & 99.2 & - & 70 & 275 & $-$ & $-$ \\
			0 & pilot study & 50 & 72.4 & - & 70 & 77.96 & 59 & 74.25  \\
			0 & surrogate & 50 & 92.8 & - & 70 & 158.91 & 148 & 65.31 \\ \hline
			0.5 & known & 50 & 100 & - & 35 & 228 & $-$ & $-$ \\
			0.5 & pilot study & 50 & 80.0 & - & 35 & 65.14 & 46 & 79.77 \\
			0.5 & surrogate & 50 & 98.2 & - & 35 & 151.95 & 131 & 78.04 \\ \hline
			1 & known & 50 & 100 & - & 47 & 296 & $-$ & $-$ \\
			1 & pilot study & 50 & 89.0 & - & 47 & 77.77 & 56.5 & 72.61 \\
			1 & surrogate & 50 & 99.6 & - & 47 & 194.20 & 166 & 281.04  \\
			\hline 
			2 & known & 50 & 100 & - & 103 & 407 & $-$ & $-$ \\
			2 & pilot study & 50 & 99.4 & - & 103 & 139.88 & 118.5 & 90.88 \\
			2 & surrogate & 50 & 100 & - & 103 & 252.01 & 227.5  & 110.40 \\
			\hline
		\end{tabular}
	\end{center}
	\label{quadTablePOW}
\end{table} 

\begin{table}[ht]
	\caption{
		Estimated concentration (OPT) under a model which violated the normality assumptions using the normality-based sample size procedure at a nominal level of 90.  To form a baseline for
		comparison, $\widehat{n}^{\mathrm{fixed}}$, shows the
		required sample size to compare the optimal embedded regime
		with standard of care. }
	\begin{center}
		\begin{tabular}{| c c c c c c c c c |} \hline
			$\Delta$ & Method for $\sigma^*$ & $n_0$ & (POW) & (OPT)
			& $\hat{n}^{\mathrm{fixed}}$ & $\mathbb{E}\hat{n}$ & $\mathrm{Med}(\hat{n})$ & $\mathrm{SD}\hat{n}$ \\ \hline
			\hline  
			0 & known & 50 & - & 100 & 70 & 586 & $-$ & $-$ \\
			0 & pilot study & 50 & - & 89.4 & 70 & 161.96 & 122 & 147.53  \\
			0 & surrogate & 50 & - & 100  & 70 & 339.32 & 317.5 & 134.60 \\ \hline
			0.5 & known & 50 & - & 100 & 35 & 485 & $-$ & $-$ \\
			0.5 & pilot study & 50 & - & 85.6 & 35 & 119.20 & 87 & 118.01 \\
			0.5 & surrogate & 50 & - & 100 & 35 & 322.16 & 285.5 & 190.63 \\ \hline
			1 & known & 50 & - & 100 & 47 & 630 & $-$ & $-$ \\
			1 & pilot study & 50 & - & 0.88 & 47 & 186.41 & 142.5 & 159.20 \\
			1 & surrogate & 50 & - & 0.99 & 47 & 390.41 & 344.5 & 191.12  \\
			\hline 
			2 & known & 50 & - & 100 & 103 & 866 & $-$ & $-$ \\
			2 & pilot study & 50 & - & 99.6 & 103 & 296.34 & 241 & 211.56 \\
			2 & surrogate & 50 & - & 100 & 103 & 537.83 & 490.5 & 277.45 \\
			\hline
		\end{tabular}
	\end{center}
	\label{quadTableOPT}
\end{table}


We also applied the projection-based sample size procedure to the two
classes of generative models described above.  Each Monte Carlo
replication consists of the following steps.   
We first
generate a pilot study of size $n_0$. The bootstrap method described
in Section 3.2 is used to calculate the minimum sample size
$\hat{n}(\mathcal{D}_{n_0})$ to achieve power
$(1 - \gamma)\times 100\%$ using 100 bootstrap replications across a grid
of potential sample sizes and then using nonlinear least squares 
to regress the estimated power on the sample sizes. 
Let $\gamma = 0.1$, $\vartheta_1 = 0.01$, and
$\vartheta_2 = 0.04$. Which corresponds to 90\% power
for a test with 5\% significance level based on a confidence interval
for $V(\pmb{\pi}^{\mathrm{opt}})$ that is being constructed using a 99\%
confidence set for $(\mu_1^*, \mu_2^*)$ and a 96\% interval for
$V(\mu_1, \mu_2)$ for each fixed value of $(\mu_1, \mu_2)$. Table
\ref{bootNormTablePOW} displays the results under the normal generative
model when sizing for condition (POW); in some cases the pilot study shows no benefit to tailoring
treatment, i.e., $\widehat{V}_{n_0} \le B_0$, in which case 
$\widehat{n}(\mathcal{D}_0) = +\infty$. Table \ref{bootNormTableOPT} displays the results
when sizing for condition (OPT) under the normal generative model.

Table \ref{bootQuadTablePOW} show the results for when the
projection-based method when sizing for condition (POW) is applied 
to the data generating model for
which the normality assumptions do not hold. The results of sizing for condition 
(OPT) for the model which the normality assumptions do not hold is contained in Table 
\ref{bootQuadTableOPT}. It can be seen that
for $\Delta \ge 0.50$ the proposed procedure attains nominal
power for (POW) for both generative models.


\begin{table}[ht]
	\caption{
		Estimated power (POW) under a model for which the normality
		assumptions hold using the projection-based sample
		size procedure at a nominal level of 90.  To form a baseline for
		comparison, $\widehat{n}^{\mathrm{fixed}}$, shows the
		required sample size to compare the optimal embedded regime
		with standard of care. }
	\begin{center}
		\begin{tabular}{| c c c c c c c c c|} \hline
			$\Delta$ & $n_0$ & (POW) & (OPT) & $\hat{n}^{\mathrm{fixed}}$
			& $\mathbb{E}\hat{n}(\mathcal{D}_{n_0})$ & $\mathrm{Med}\{\hat{n}(\mathcal{D}_{n_0})\} $ &
			$\mathrm{SD}\hat{n}(\mathcal{D}_{n_0}) $ & 
			$P\left\lbrace \hat{n}(\mathcal{D}_{0}) = \infty\right\rbrace$ \\ \hline
			
			0 & 50 & 85.17 & - & 74 & 381.94 & 342.5 & 112.46 & 0.42 \\ 
			0.5 & 50 & 99.13 & - & 111 & 316.01 & 292.5 & 83.89 & 0.21 \\ 
			1 & 50 & 99.73 & - & 151 & 312.39 & 299 & 71.96 & 0.17 \\ 
			2 & 50 & 100 & - & 251 & 364.95 & 361.5 & 62.78 & 0.06 \\ 
			\hline  
		\end{tabular}
	\end{center}
	\label{bootNormTablePOW}
\end{table}

\begin{table}[ht]
	\caption{
		Estimated concentration (OPT) under a model for which the normality
		assumptions hold using the projection-based sample
		size procedure at a nominal level of 90.  To form a baseline for
		comparison, $\widehat{n}^{\mathrm{fixed}}$, shows the
		required sample size to compare the optimal embedded regime
		with standard of care. }
	\begin{center}
		\begin{tabular}{| c c c c c c c c |} \hline
			$\Delta$ & $n_0$ & (POW) & (OPT) & $\hat{n}^{\mathrm{fixed}}$
			& $\mathbb{E}\hat{n}(\mathcal{D}_{n_0})$ & $\mathrm{Med}\{\hat{n}(\mathcal{D}_{n_0})\} $ &
			$\mathrm{SD}\hat{n}(\mathcal{D}_{n_0}) $ \\ \hline
			0 & 50 & - & 100 & 74  & 114.73 & 113 & 9.72 \\ 
			0.5 & 50 & - & 100 & 111 & 107 & 106 & 9.59 \\ 
			1 & 50 & - & 99.8 & 151 & 112.22 & 111 & 10.69 \\ 
			2 & 50 & - & 99.4 & 251 & 124.3 & 123 & 13.22 \\ 
			\hline  
		\end{tabular}
	\end{center}
	\label{bootNormTableOPT}
\end{table}


\begin{table}[ht]
	\caption{
		Estimated power (POW) under a model which violated the normality assumptions using the projection-based sample size procedure at a nominal level of 90.  To form a baseline for
		comparison, $\widehat{n}^{\mathrm{fixed}}$, shows the
		required sample size to compare the optimal embedded regime
		with standard of care. }
	\begin{center}
		\begin{tabular}{| c c c c c c c c c  |} \hline
			$\Delta$ & $n_0$ & (POW) & (OPT) & $\hat{n}^{\mathrm{fixed}}$ 
			& $\mathbb{E}\hat{n}(\mathcal{D}_{n_0})$ & $\mathrm{Med}\{\hat{n}(\mathcal{D}_{n_0})\} $ & $\mathrm{SD}\hat{n}(\mathcal{D}_{n_0})$ & 
			$P\left\lbrace \hat{n}(\mathcal{D}_0) = \infty\right\rbrace$ \\ \hline	
		0 & 50 & 84.28 & - & 70 & 482.18 & 401 & 250.04 & 0.34 \\ 
		0.5 & 50 & 92.96 & - & 35 & 527.52 & 491 & 208.04 & 0.41 \\ 
		1 & 50 & 98.55 & - & 47 & 556.37 & 507.5 & 227.93 & 0.28 \\ 
		2 & 50 & 99.49 & - & 103 & 594.68 & 536 & 281.35 & 0.14 \\ 
			\hline  
		\end{tabular}
	\end{center}
	\label{bootQuadTablePOW}
\end{table}

\begin{table}[ht]
	\caption{
		Estimated concentration (OPT) under a model which violated the normality assumptions using the projection-based sample size procedure at a nominal level of 90.  To form a baseline for
		comparison, $\widehat{n}^{\mathrm{fixed}}$, shows the
		required sample size to compare the optimal embedded regime
		with standard of care. }
	\begin{center}
		\begin{tabular}{| c c c c c c c c |} \hline
			$\Delta$ & $n_0$ & (POW) & (OPT) & $\hat{n}^{\mathrm{fixed}}$ 
			& $\mathbb{E}\hat{n}(\mathcal{D}_{n_0})$ & $\mathrm{Med}\{\hat{n}(\mathcal{D}_{n_0})\} $ & $\mathrm{SD}\hat{n}(\mathcal{D}_{n_0})$ \\ \hline	
			0 & 50 & - & 96.6 & 70 & 84.13 & 83 & 11.66 \\ 
			0.5 & 50 & - & 93.6 &  35 & 88.93 & 88 & 12.08  \\ 
			1 & 50 & - & 95.6 & 47 & 90.29 & 88 & 14.16 \\ 
			2 & 50 & - & 97.6 & 103 & 90.53 & 88 & 22.73 \\ 
			\hline  
		\end{tabular}
	\end{center}
	\label{bootQuadTableOPT}
\end{table}



\section{Discussion} 
We proposed two sample size procedures for two-stage SMARTs when the
objective is estimation and evaluation of an optimal dynamic treatment
regime.  These procedurs can be used to design SMARTs or conduct power
analyses for observational studies.  Furthermore, a comparison of the
sample size required for construction of a high-quality estimator of
an optimal treatment regime with the sample size required for
comparison of fixed treatment sequences (or another simple comparison
commonly used to size a SMART) can generate new insights into the cost
of precision medicine in a given problem domain.

The proposed procedures were developed under two extremes
in terms of the structure imposed on the underlying generative
model. At one extreme, we assumed 
correctly specified parametric models for several functionals
of the generative model including the optimal regime; and, at the other
extreme, we only imposed moment
conditions on a possibly misspecified analysis model.  There is large
class of intermediate models that could be constructed from these
two base approaches.   Furthermore, while the proposed approaches
focused
on regression-based estimators they can be extended to
classification-based or direct-search estimators \citep[][]{Orellana10,
  baqun, yingqi, zhang2012estimating,baqun2, yingqi2,
  zhou2017residual, yingqi3, midas} which are becoming increasingly
popular; we leave the details of this extension to future work.


\section{Acknowledgments}
This work was supported by the National Science Foundation
(DMS 1513579, 1557733, 1555141) and the National
Institutes of Health (P01 CA142538). 

\bibliographystyle{Chicago}
\bibliography{sampleSize}

\end{document}